\documentclass[5p,sort&compress]{elsarticle}
\usepackage[english]{babel}
\usepackage{subcaption} 
\usepackage{wrapfig}
\usepackage{graphicx}
\usepackage{epstopdf}
\usepackage{multicol}
\usepackage{tabularx}
\usepackage{amssymb,amsmath}
\usepackage{mathrsfs}
\usepackage{xfrac}
\usepackage{textcomp}
\usepackage{bm}
\usepackage{color}
\usepackage{ulem}
\usepackage{setspace} %Задание интервала с возможностью изменения 
\usepackage{float}
\definecolor{dblue}{rgb}{0,0,0.8}

\definecolor{dred}{rgb}{0.7,0,0}

\definecolor{dgreen}{rgb}{0.,0.6,0}

\makeatletter % \lambdabar Shamelessly stolen from revtex.sty
\def\lambdabar{\protect\@lambdabar}
\def\@lambdabar{%
\relax
\bgroup
\def\@tempa{\hbox{\raise.73\ht0
\hbox to0pt{\kern.25\wd0\vrule width.5\wd0
height.1pt depth.1pt\hss}\box0}}%
\mathchoice{\setbox0\hbox{$\displaystyle\lambda$}\@tempa}%
{\setbox0\hbox{$\textstyle\lambda$}\@tempa}%
{\setbox0\hbox{$\scriptstyle\lambda$}\@tempa}%
{\setbox0\hbox{$\scriptscriptstyle\lambda$}\@tempa}%
\egroup
}
\makeatother% end of \lambdabar

\usepackage{hyperref}
\journal{}
\begin{document}

\begin{frontmatter}

\title{Scattering of ultrarelativistic electrons in ultrathin crystals
%\tnoteref{mytitlenote}
}

\author[addr_KIPT,addr_KhUni]{N.F.~Shul'ga}\corref{mycorrespondingauthor}
%\ead{shulga@kipt.kharkov.ua}
\author[addr_KIPT]{S.N.~Shulga}
%\ead{shulga.serge@gmail.com}
%\author[addr_KIPT]{S.N.~Shulga}
\cortext[mycorrespondingauthor]{Corresponding author, e-mail: shulga@kipt.kharkov.ua}
\address[addr_KIPT]{National Science Center "Kharkov Institute of Physics and Technology". 1, Akademichna str. Kharkiv 61108 Ukraine}
\address[addr_KhUni]{Karazin Kharkiv National University. 4, Svobody sq. Kharkiv 61000 Ukraine}

\begin{abstract}
Quantum theory is proposed of high energy electrons scattering in ultrathin crystals. This theory is based upon a special representation of the scattering amplitude in the form of the integral over the surface surrounding the crystal, and on the spectral method of determination of the wave function. The comparison is performed of quantum and classical differential scattering cross-sections in the transitional range of crystal thicknesses, from those at which the channeling phenomenon is not developed up to those at which it is realized. It is shown that in this thickness range the quantum scattering cross-section, unlike the classical one, contains sharp peaks corresponding to some specific scattering angles, that is connected with the diffraction of the incident plane wave onto the periodically distributed crystal atomic strings. It is shown that the value of the scattering cross-section in the peaks varies periodically with the change of the target thickness. We note that this must lead to a new interference effect in radiation that is connected with the rearrangement of incident wave packet in transitional area of crystal thicknesses.
\end{abstract}

\begin{keyword}
Keywords: ultrathin crystal, scattering, channeling, spectral method,  electron diffraction.\\
%\texttt{elsarticle.cls}\sep \LaTeX\sep Elsevier \sep template
%\MSC[2010] 00-01\sep  99-00\\
%PACS numbers: 29.27.-a, 61.85.+p, 34.80.Pa, 61.05.J
\end{keyword}

\end{frontmatter}

%\linenumbers

\section{\label{sec:level1}Introduction}

At passing of high energy charged particles through crystals the phenomenon of channeling is possible, at which the particles move inside the channels formed by crystal atomic strings or atomic planes, being periodically deviated to small angles from the channel direction \cite{Lindh65,AkhiShul96}. In ultrathin crystals there is no room for channeling phenomenon to be developed (see {\it Fig.\;1}). However, there remains the possibility of appearance of several coherence and interference effects at interaction of particles with crystal atoms (at high energies the attention to this fact was paid in the works \cite{Ferret50,TerMik53,Uber56}). Such phenomena take place in several electromagnetic processes at high energies in crystals, such as scattering, radiation and electron-positron pairs creation (see \cite{AkhiShul96,TerMik72} and references therein). 

The present work is devoted to the development of classical and quantum theories of high energy charged particles scattering in transitional range of crystal thicknesses, from those at which the channeling phenomenon is not developed, up to those at which this phenomenon is realized. Quantum theory is based upon the special representation of the scattering amplitude \cite{BonShul98} in the form of the integral over a surface surrounding the region of influence of external crystal field onto the particle (in the considered problem this corresponds to the field of entire crystal), and upon the development of numeric methods of calculating of the wave function inside the crystal, that is realized by using the so-called spectral method of solving wave equations \cite{FeitFleck82,DabagOgn88,ShulSysch16}. The classical theory is based upon the solution of the particle motion equation by numerical methods \cite{Hamm62}. The main attention was paid to the comparative analysis of quantum and classical characteristics of the scattering process at different crystal thicknesses and particle energies.
\begin{figure}
\includegraphics[width=0.57\columnwidth]{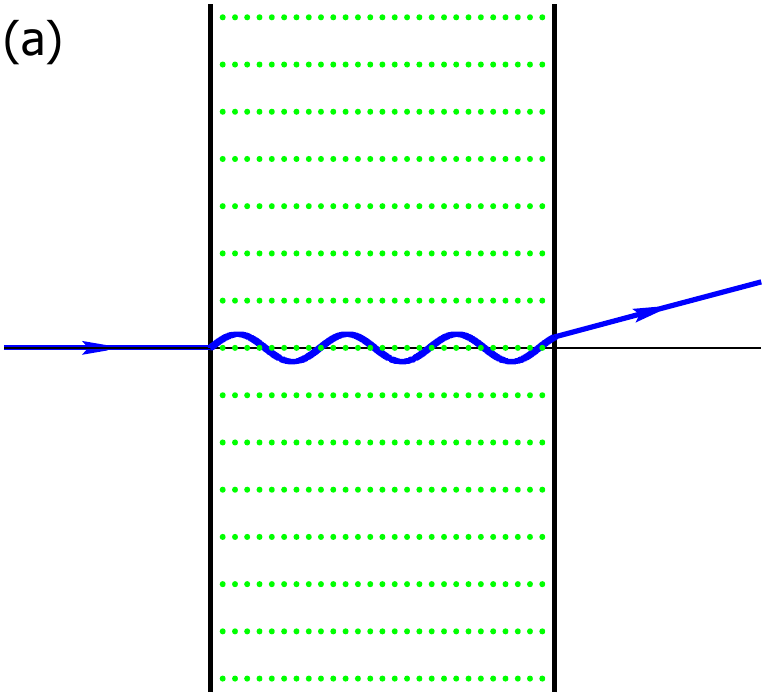}\quad
\includegraphics[width=0.344\columnwidth]{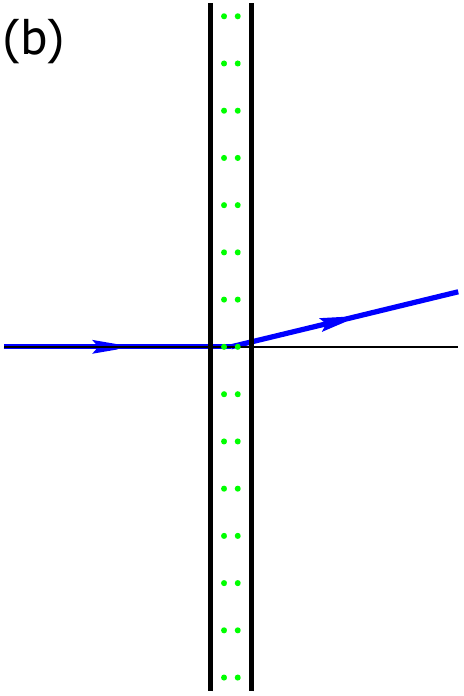}
\caption{\label{fig1}Scattering in channeling regime (a) and in absence of channeling (b).}
\end{figure}

In the last years new possibilities have opened to carry out experimental research in this field, that is connected with the development of technologies of ultrathin crystals production for the accelerator experiments, and improving of the beam parameters \cite{Guidi12,MothBrees12}. Due to this, one can speak now about discovering quantum and classical effects in scattering discussed in the present paper.

\section{\label{sec:level1}Quantum Theory}

Let us consider the scattering of relativistic electrons incident onto a thin crystal along one of its crystal axes. Differential scattering cross-section and scattering amplitude in this case are defined by the following formulas \cite{AkhiShul96}:
\begin{equation}
\frac{d\sigma}{do} = {\left|{a\left( \vartheta  \right)}\right|^2},
\end{equation}
\begin{equation}
a\left( \vartheta  \right) =  - \,\,\frac{1}{{4\pi {\hbar ^2}}}\int {{d^3}r\,{e^{ - \,\frac{i}{\hbar }{\bf{p'r}}}}\overline{u'} {\gamma _0}U\left( {\bf{r}} \right)\psi \left( {\bf{r}} \right)}
\label{eq2}.
\end{equation}
where $\vartheta$ is the scattering angle, $\psi(\bf{r})$ -- the wave function of the electron passing through the crystal, $\overline{u'}$  and $\bf{p'}$ -- respectively bispinor and momentum of the scattered electron, and $U(\bf{r})$ -- the potential energy of interaction of the electron with the crystal lattice field (we use the system of units in which the light velocity is equal to one, $c=1$). 

At falling of fast particles onto the crystal along one of its axes (we will name it $z$ axis) the correlations between consequent collisions of the particle with the lattice atoms are substantial. As a result of these correlations, the particle motion is defined mainly by the continuous potential of crystal atomic strings situated parallel to the $z$ axis, being the lattice potential averaged along this axis \cite{Lindh65,AkhiShul96}: 
\begin{equation}
{U_c}\left( {{\bm{\rho }},z} \right) = \frac{1}{L}\int\limits_0^L {dz\,U\left( {\bf{r}} \right)},
\label{eq3}
\end{equation}
where $L$ is the crystal thickness and $\bm \rho$ -- the co-ordinates $(x,y)$ in the plane orthogonal to the $z$ axis.

The crystal lattice potential is localized in the region of the external field action onto the particle. The scattering amplitude (\ref{eq2}) in this case, with use of the Gauss theorem, can be presented in the form of integral over a closed surface surrounding the external field region \cite{BonShul98}:
\begin{equation}
a\left( \vartheta  \right) =  - \,\,\frac{i}{{4\pi \hbar }}\oint {d{\bf{S}}\,{e^{ - \,\frac{i}{\hbar }{\bf{p'r}}}}\,\overline {u'} \,{\bf{\gamma }}\,\psi \left( {\bf{r}} \right)},
\label{eq4}
\end{equation}
where $d\bf{S}$ is the element of the surface surrounding the crystal. It is also substantial that the surface integral in (\ref{eq4}) does not depend on the surface form, and the only requirement here is that this surface surround the entire area of the external field action. In the considered problem it is convenient to choose as such a closed surface a cylinder whose bases coincide with the crystal sides. By neglecting the contribution of the cylinder lateral side parallel to the $z$ axis in the surface integral (\ref{eq4}), we come to the following expression for the scattering amplitude (\ref{eq4}):
\begin{equation}
a\left( \vartheta  \right) =  - \frac{i}{{4\pi \hbar }}\int {\left. {{d^2}\rho \,{e^{ - \,\frac{i}{\hbar }{\bf{p'r}}}}\,\overline {u'} \,{\gamma _z}\,\psi \left( {\bf{r}} \right)} \right|} _{z = 0}^{z = L}.
\label{eq5}
\end{equation}

Before entering into the crystal, the electron wave function is represented as the plane wave ${\left. {\psi \left( {\bf{r}} \right)} \right|_{z < 0}} = {u_p}\exp \left( {\frac{i}{\hbar }{\bf{pr}}} \right)
$ defined by the initial momentum  $\bf{p}$ and bispinor $u_p$. Due to this, the contribution of the term with $z=0$ into the amplitude (\ref{eq5}) for non-zero scattering angles disappears. Here, by detaching in the wave function $\psi({\bf{r}})$ the plane wave factor of the particle falling onto the crystal $\psi \left( {\bf{r}} \right) = \Phi \left( {\bf{r}} \right)\exp \left( {\frac{i}{\hbar }{\bf{pr}}} \right)
$, we write down the scattering amplitude in the form:
\begin{equation}
a\left( \vartheta  \right) =  - \frac{i}{{4\pi \hbar }}\int {\left. {{d^2}\rho \,{e^{\frac{i}{\hbar }{\bf{qr}}}}\,\overline {u'} \,{\gamma _z}\,\Phi \left( {\bf{r}} \right)} \right|}_{z=L},
\label{eq6}
\end{equation}
where $\bf{q}=\bf{p}-\bf{p'}$ is the momentum transmitted to the crystal at scattering.

The electron wave function in the external field $U({\bf{r}})$ is a solution of the squared Dirac equation \cite{AkhiShul96}
\begin{equation}
\left[ {{{\left( {\varepsilon  - U\left( {\bf{r}} \right)} \right)}^2} - {{\left( {i\hbar \nabla } \right)}^2} - {m^2} + i\hbar {\bf{\alpha }}\nabla U\left( {\bf{r}} \right)} \right]\psi \left( {\bf{r}} \right) = 0,
\end{equation}
where $m$ and $\varepsilon$ are the particle mass and energy respectively, and ${\bm{\alpha}} = {\gamma _0}{\bm{\gamma}}
$.

For $\Phi \left( {\bf{r}} \right)$ we come to the following equation: 
\begin{equation}
i\hbar v{\partial _z}\Phi  = \left[ {\frac{{{{{\bf{\hat p}}}^2}}}{{2\varepsilon }} + U\left( {\bf{r}} \right) - \frac{{i\hbar }}{{2\varepsilon }}{\bf{\alpha }}\nabla U\left( {\bf{r}} \right) - \frac{{{U^2}}}{{2\varepsilon }}} \right]\Phi,
\label{eq8}
\end{equation}
where $v = {p \mathord{\left/{\vphantom {p \varepsilon }} \right.
 \kern-\nulldelimiterspace} \varepsilon }$ is the particle velocity, ${\bf{\hat p}} =  - i\hbar \nabla$ is the momentum operator and the $z$ axis is directed along $\bf{p}$.

Characteristic values for the scattering angles of high energy electrons in thin crystal are small compared to one. In this case, by resolving the Eq.\;(\ref{eq8}), we can neglect the spin effects in scattering (spin-field interaction) and the terms proportional to ${{{U^2}} \mathord{\left/
 {\vphantom {{{U^2}} {2\varepsilon }}} \right.\kern-\nulldelimiterspace} {2\varepsilon }}$ and ${{\hat p_z^2} \mathord{\left/{\vphantom {{\hat p_z^2} {2\varepsilon }}} \right.\kern-\nulldelimiterspace} {2\varepsilon }}$. So, by selecting the plane wave bispinor ${u_0}$ in the function $\Phi \left( {\bf{r}} \right)$, 
\begin{equation}
\Phi \left( {\bf{r}} \right) = \varphi \left( {\bf{r}} \right){u_0},
\end{equation}
we get the following equation for $\varphi \left( {\bf{r}} \right)$:
\begin{equation}
i\hbar v{\partial _z}\varphi  = \left[ {\frac{{{\bf{\hat p}}_ \bot ^2}}{{2\varepsilon }} + U\left( {\bf{r}} \right)} \right]\varphi,
\label{eq10}
\end{equation}
where ${{\bf{\hat p}}_ \bot } =  - i\hbar \frac{\partial }{{\partial {\bf{\rho }}}}$.

The solution of (\ref{eq10}) inside an ultrathin crystal can be found with use of numerical methods based upon the spectral method that has been used for determining eigenstates in the complicated configuration fields \cite{FeitFleck82,DabagOgn88,ShulSysch16}. By using this solution for the wave function $\varphi \left( {{\bm{\rho }},z} \right)$ at $z=L$, the scattering cross-section and amplitude in the region of small scattering angles can be written in the form:
\begin{equation}
\frac{{{d^2}\sigma }}{{{d^2}\vartheta }} = {\left| {{a_L}\left( {{{\bf{q}}_ \bot }} \right)} \right|^2},
\label{eq11}
\end{equation}
\begin{equation}
{a_L}\left( {{{\bf{q}}_ \bot }} \right) =  - \frac{{ip}}{{2\pi \hbar }}\int {{d^2}\rho \,{e^{\frac{i}{\hbar }{{\bf{q}}_ \bot }{\bf{\rho }}}}\varphi \left( {{\bf{\rho }},L} \right)},
\end{equation}
where ${{\bf{q}}_ \bot } = p\bm\vartheta$ is the transversal component of the momentum transmitted to the crystal and $\bm{\vartheta}  = \left( {{\vartheta _x},{\vartheta _y}} \right)$ is the scattering angle $\left( {\vartheta \ll 1} \right)$. Here we used the facts that at small scattering angles $\overline{u'} {\gamma _z}{u_p} \approx 2{p_z} \approx 2p$ and that the factor $\exp \left( {i{q_z}L} \right)$ in (\ref{eq6}) disappears in the scattering cross-section.

\section{\label{sec:level1}Classical Theory}

Now let us consider the fast electrons scattering in thin crystal on the basis of the classical mechanics. The particle motion in this case is defined by its classical trajectory, being the solution of classical equations of motion, and satisfies to the given initial conditions.

The differential scattering cross-section in classical mechanics corresponds to the elementary surface in the impact parameter space, from which the particle is scattered into the elementary solid angle $do \approx {d^2}\vartheta$:
\begin{equation}
d{\sigma _{cl}} = {d^2}b\left( \bm\vartheta  \right).
\label{eq13}
\end{equation}
For calculating this cross-section we need to find the deflection function of the particle in the external field $\bm\vartheta  = \bm\vartheta \left( {\bf{b}} \right)$, that is the dependence of the particle scattering angle $\bm\vartheta  = \left( {{\vartheta _x},{\vartheta _y}} \right)$ on the impact parameter ${\bf{b}} = \left( {{b_x},{b_y}} \right)$, and then to perform the inversion of this function, {\it i.e.} to define the dependence ${\bf{b}} = {\bf{b}}\left( \bm\vartheta  \right)$ (see, {\it e.g.} \cite{AkhiShul96,Newt66}). In a complex field, as inside the crystal, the deflection function $\bm\vartheta  = \bm\vartheta \left( {\bf{b}} \right)$ is quite a complicated function of the co-ordinates $b_x$ and $b_y$. It is substantial that the deflection function inversion is not single-valued in the common case. With taking into account this ambiguity, the classical scattering cross-section (\ref{eq13}) can be presented in the following form:
\begin{equation}
d\sigma \left( \bm\vartheta  \right) = \sum\limits_n {{d^2}{b_n}\left( \bm\vartheta  \right)}  = \sum\limits_n {{{\left. {\frac{1}{{{{\left| {{{\partial \bm\vartheta } \mathord{\left/
 {\vphantom {{\partial \bm\vartheta } {\partial {\bf{b}}}}} \right.
 \kern-\nulldelimiterspace} {\partial {\bf{b}}}}} \right|}_n}}}} \right|}_{{\bf{b}} = {{\bf{b}}_n}\left( \bm\vartheta  \right)}}{d^2}\vartheta },
\label{eq14}
\end{equation}
where $\left| \partial\bm\vartheta/\partial\bf{b}\right| = \left|\partial(\vartheta_x,\vartheta_y)/\partial(b_x,b_y)\right|$ is the transition determinant from the variables ${\vartheta _x} = {\vartheta _x}\left( {{b_x},{b_y}} \right)$  and ${\vartheta _y} = {\vartheta _y}\left( {{b_x},{b_y}} \right)$ to $b_x$ and $b_y$, with the subsequent inversion of the deflection function. The summation in (\ref{eq14}) is performed over the single-valued branches of the deflection function. 

The formula (\ref{eq14}) for the scattering cross-section can also be written in the form:
\begin{equation}
\frac{{{d^2}{\sigma _{cl}}\left( \bm\vartheta  \right)}}{{{d^2}\vartheta }} = \int {{d^2}b\,\,\delta \left( {\bm\vartheta  - \bm\vartheta \left( {\bf{b}} \right)} \right)}.
\end{equation}
Let us note that this expression for the classical scattering cross-section can be easily obtained from the quantum one (\ref{eq11}) in the frames of the quasi-classical approximation of quantum mechanics, in the case of a single-valued correspondence between the particle's scattering angle and its impact parameter. 

In the common case the deflection function can be expressed via classical particle trajectories after their exit from the external field area (in our case from the crystal). At small scattering angles this dependence is defined by the relation
\begin{equation}
\bm\vartheta \left( {{\bf{b}},L} \right) = \frac{1}{v}{v_ \bot }\left( {{\bf{b}},L} \right),
\end{equation}
where ${v_ \bot }\left( {{\bf{b}},L} \right)$ is the transversal component of the particle velocity at $z=L$. The velocity ${v_ \bot }\left( {{\bf{b}},z} \right)$ is defined as the solution of the classical equation of motion. At the particle motion in the continuous strings potential field (\ref{eq3}), this equation, with the precision up to the terms of the order of $\mathcal{O}\!\left( {{{v_ \bot ^2} \mathord{\left/ {\vphantom {{v_ \bot ^2} {{v^2}}}} \right.
 \kern-\nulldelimiterspace} {{v^2}}}} \right)$
 has the following form \cite{Lindh65,AkhiShul96}:
\begin{equation}
\ddot \rho  =  - \frac{1}{\varepsilon }\frac{\partial }{{\partial \rho }}U\left( {\rho ,z \approx vt} \right).
\label{eq17}
\end{equation}
The solution of (\ref{eq17}) can be found on the basis of numerical methods (see, {\it e.g.}, \cite{Hamm62}) for a large number of particles falling onto a crystal at different values of impact parameters. This lets one develop the procedure of numerical calculation of the scattering cross-section for a particle beam falling onto a crystal with random uniformly distributed impact parameters. At this, the probability of the particles scattering into the solid angle interval $\left( {\bm\vartheta ,\bm\vartheta  + d\bm\vartheta } \right)$ is defined by the relation of particle number $dN\!\left( \bm\vartheta  \right)
$ that exited into this interval, to the full number of incident particles $N$. The scattering cross-section is connected with $dN$ by the relation 
\begin{equation}
dN(\bm\vartheta)=\frac{N}{S}d{\sigma_{cl}}(\bm\vartheta),
\end{equation}
where $S$ is the transverse size of the crystal surface on which the particles fall down.

\section{\label{sec:level1}Comparative analysis of quantum and classical effects in scattering}

On the basis of the above-stated methods, one can carry out numerical calculations of quantum and classical elastic scattering cross-sections of relativistic charged particles in thin crystals, and perform the comparative analysis of the quantum and classical effects in scattering. We will present some results of calculations for relativistic electrons with different energies in ultrathin Si crystals that are oriented by their $\left\langle {100} \right\rangle$ axis to the incident beam. The continuous potential of the whole ensemble of atomic strings represents itself quite a complicated two-dimensional periodical function of the co-ordinates $\left( {x,y} \right)$ in the plane orthogonal to the $z$ axis (see \cite[Fig.\;6.16]{AkhiShul96}). The calculation of quantum and classical scattering cross-sections in such field can only be performed on the basis of numerical methods. 

In quantum calculations, in quality of the initial state $\varphi \left( {{\bf{\rho }},z = 0} \right)$ we used a wide wave packet covering a large number of crystal atomic strings whose axes are periodically situated in the transverse plane.

The calculations based upon the classical theory were performed for $10^7$ particles falling onto the crystal with the impact parameters randomly distributed over the elementary cell in its transverse plane. 

\begin{figure}
\includegraphics[width=1.0\columnwidth]{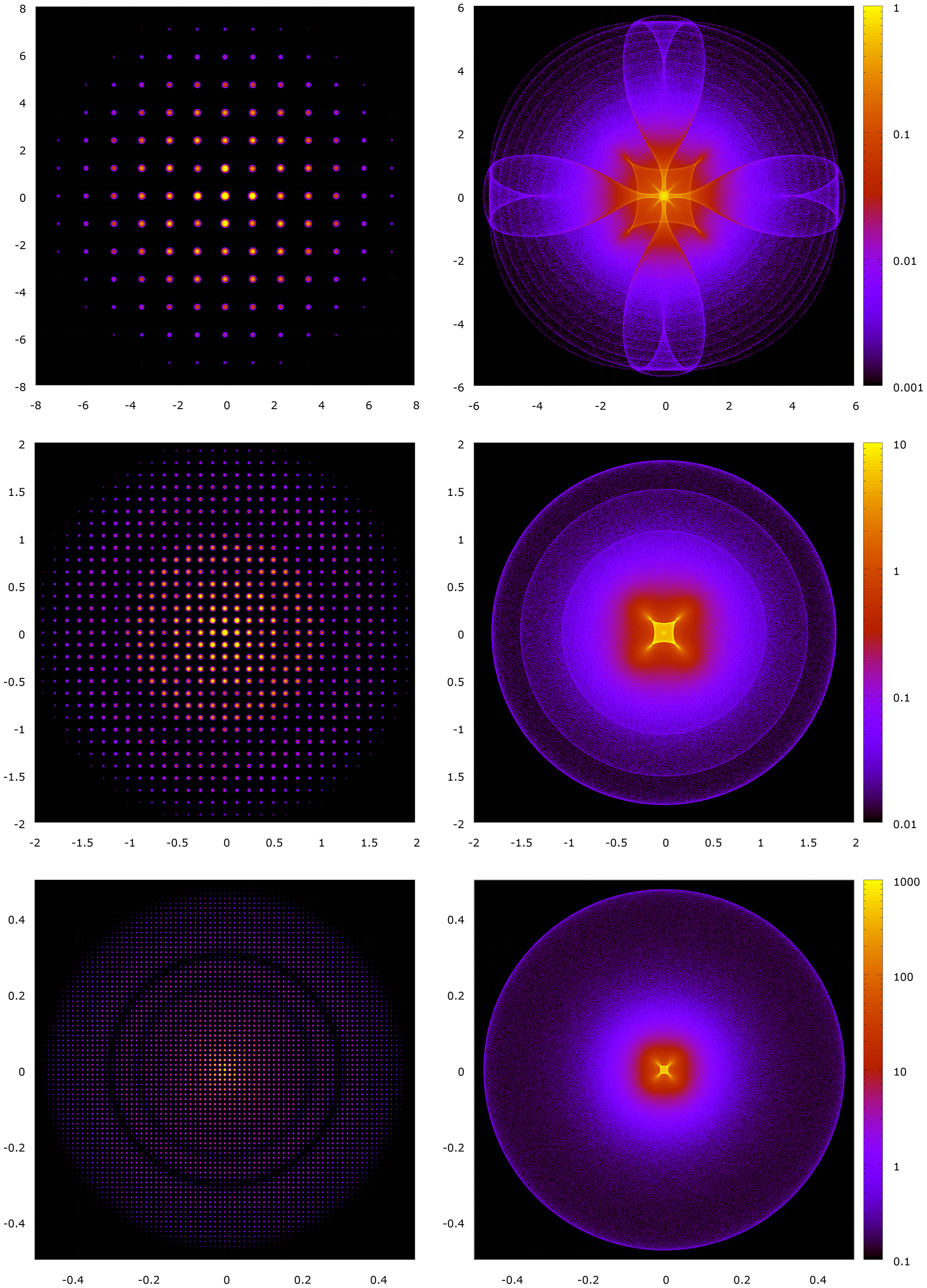}
\caption{\label{fig2}Simulation of quantum (left) and classical (right) scattering pictures of electrons in 1000Å Si $\langle100\rangle$ crystal. Kinetic energies of particles are 5 MeV, 50 MeV and 500 MeV, from top to bottom.}
\end{figure}

In Fig.\;2 and Fig.\;3 we present the results of calculations of quantum and classical (Fig.\;2) and quantum (Fig.\;3) scattering cross-sections for ultrarelativistic electrons in Si crystals in the transitional thickness range, at which the channeling phenomenon is still not realized.

The results obtained testify that the picture of angular distributions of the particles scattered by crystal is quite complicated, being substantially depending on both crystal thickness and particle energy. Classical scattering cross-sections contain sharp edges between the permitted and forbidden scattering angles areas. The presence of such edges and the substantial increase of the scattering cross-section on them are typical to the rainbow scattering phenomenon of the particles in external field, that is explained by turning into zero of the determinant $\left| {{{\partial \left( {{\vartheta _x},{\vartheta _y}} \right)} \mathord{\left/{\vphantom {{\partial \left( {{\vartheta _x},{\vartheta _y}} \right)} {\partial \left( {{b_x},{b_y}} \right)}}} \right.\kern-\nulldelimiterspace} {\partial \left( {{b_x},{b_y}} \right)}}} \right|$ in Eq.~(\ref{eq14}). Let us note that in the considered problem this determinant turns into zero not at some single values of impact parameters, as it takes place for the particle scattering in the central field \cite{AkhiShul96,Newt66}, but along some curves in the impact parameter plane. This fact opens new possibilities for the study of the rainbow scattering phenomenon manifestations. The attention to the possibility of such an effect at fast protons scattering in ultrathin crystal was paid in the work \cite{Nesk86} and ongoing works, for the case at which the particle trajectories are nearly rectilinear. The attention to the possibility of the rainbow scattering phenomenon of relativistic electrons on crystal strings was paid in the works \cite{FomShul79}. 

Unlike classical scattering cross-section, the quantum one, as our calculations show, contains sharp peaks at some values of scattering angles. These peaks are caused by the interference effect at the wide wave packet (plane wave) scattering on the crystal atomic strings that are periodically situated inside the crystal. Thanks to this interference, the transmitted momentum values at the particle scattering on crystal take the values proportional to the whole number of inverse lattice vectors $\bf g$. It is essential here that the maxima positions in the particles angular distributions depend on the character of distribution of the atomic strings in the transverse plane. In this connection, we can tell that the angular distributions analysis of the particles scattered on a thin crystal is analogical to the X-Ray analysis of the crystal structure, but with such difference that in our case we deal not with single crystal atoms but with atomic strings as elementary scattering objects. 

The distances between interference maxima in particle angular distributions are defined by the relation $\Delta \vartheta  \approx {{2\pi }\mathord{\left/{\vphantom {{2\pi } {p{a_r}}}} \right.\kern-\nulldelimiterspace}{p{a_r}}}$, where $a_r$ is the distance between the atomic strings. That's why it is necessary for the experimental discovery of such maxima that the angular divergence of the particles in the incident beam $\Delta \psi$ be small compared to the distance between these maxima, $\Delta \psi \ll \Delta \vartheta$ . For the 5\,MeV electrons falling onto the Si crystal along   axis, we have $\Delta \vartheta \sim 1 \,{\mathop{\rm mrad}\nolimits}$ (see Fig.\;2). The critical axial channeling angle in this case is about ${\psi _c} \approx 5{\mathop{\rm mrad}\nolimits}$. So, for such energies the condition $\Delta \psi \ll \Delta \vartheta$ is quite reachable in the experiment.

As the full particle energy increases, the distance between the interference maxima decreases, and the quantum scattering picture approaches to the classical one.
\begin{figure*}
\includegraphics[width=1.0\textwidth]{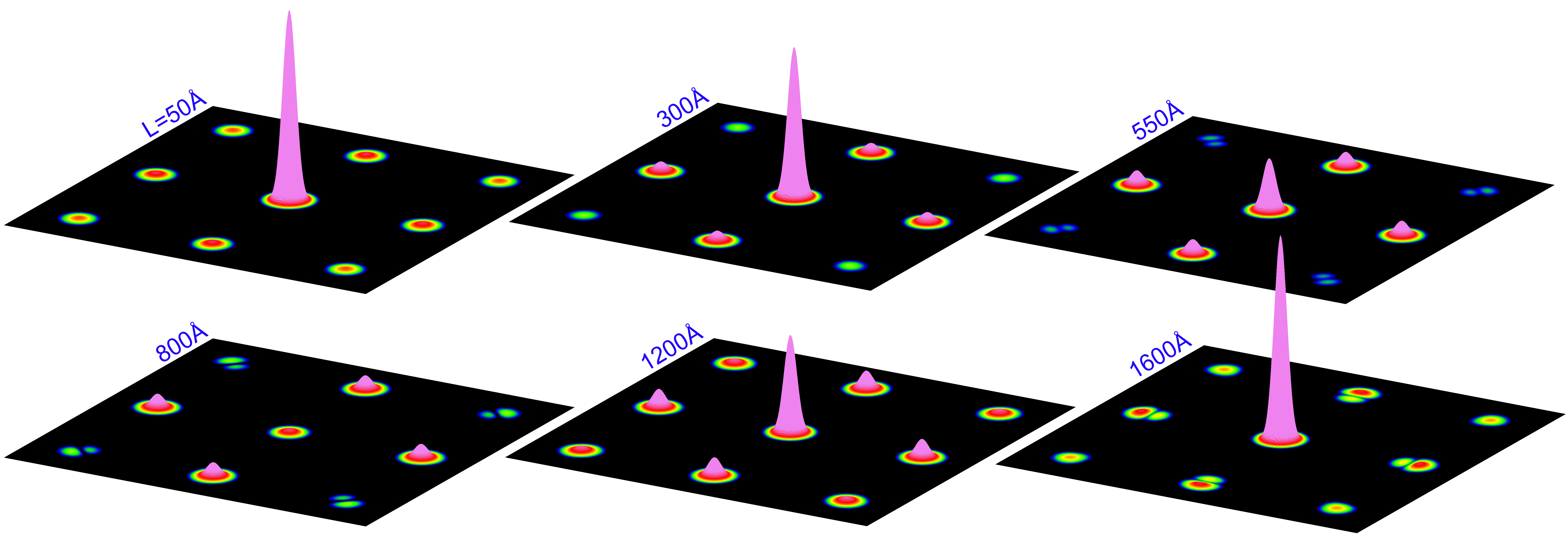}% Here is how to import EPS art
\caption{\label{fig3}Central part of the scattering picture of 5\,MeV electrons by thin Si $\langle100\rangle$ crystal of different thicknesses (indicated on figure). Visible height of visible peaks is in linear scale, color scale is logarithmic.}
\end{figure*}
Note that in the considered range of thicknesses the quantum levels of transverse motion are still not developed (for these levels to be formed it is necessary that the particle perform at least a few oscillations inside the channel). At the same time, in this thickness range there realizes a substantial rearrangement of the wave packet (plane wave) falling onto the crystal, that is connected with the periodicity of the atomic strings positions. Due to this rearrangement, the formation of interference maxima in angular distributions of scattered particles is realized. Let us also pay attention to the fact that with the target thickness increase the values of scattering cross-section in these maxima  change. There is some periodicity in these intensity  changes, the period being of the order of the length $l\sim{a/{\psi _c}}$ along the $z$ axis, that substantially exceeds the distance $a$ between the string atoms (see Fig.\;3). This, in its turn, must result in an oscillatory dependence of electron radiation characteristics in the crystal, and moreover, such an effect is possible for single electrons. The detailed analysis of such effects, however, exceeds the frames of the present work.

Note that the structure of the obtained quantum angular distributions of scattered particles is analogical to the structure of angular distributions of particles obtained in the electron microscopy (see, {\it e.g.} \cite{Hirsch65, Ohts83}).  In the considered problem, however, we deal with the particles whose energies substantially exceed those of the particles used in the electron microscopy. Moreover, in the electron microscopy the analysis of the process of particle scattering is performed on the basis of two- and many-wave formalism. In the present work such analysis is performed on the basis of the spectral method of direct numeric calculation of the wave function and the scattering amplitude. This opens new possibilities in the description of physical processes at interaction of high energy particles with crystals, such as the analysis of transition from the quantum picture of scattering of charged particles in crystal to the classical one, the rainbow scattering phenomenon, radiation in the transitional region of crystal thickness at which the channeling regime is still not formed, {\it etc}. This method can also be applied to the problems connected with the electron microscopy. 

\section{\label{sec:level1}Acknowledgements}

The work is partially supported by the grant of the NAS of Ukraine (project no. $\Phi5-2016$), by the grant of the Ministry of Education of Ukraine, (project no. 0115U000473) and by the grant of SSFR of Ukraine (project no. $\Phi64-2016$).

\nocite{*}

\bibliography{ultrathin}% Produces the bibliography via BibTeX.

\end{document}